\documentclass[12pt,preprint]{aastex}
\usepackage{amsfonts,amsmath,graphicx,natbib }

\shortauthors{Bower et al.}
\shorttitle{Variable Size of Sgr A*}
\begin{document}

\newcommand\degd{\ifmmode^{\circ}\!\!\!.\,\else$^{\circ}\!\!\!.\,$\fi}
\newcommand{\etal}{{\it et al.\ }}
\newcommand{\uv}{(u,v)}
\newcommand{\rdm}{{\rm\ rad\ m^{-2}}}
\newcommand{\msuny}{{\rm\ M_{\sun}\ y^{-1}}}
\newcommand{\mylesssim}{\stackrel{\scriptstyle <}{\scriptstyle \sim}}
\newcommand{\lsim}{\stackrel{\scriptstyle <}{\scriptstyle \sim}}
\newcommand{\gsim}{\stackrel{\scriptstyle >}{\scriptstyle \sim}}
\newcommand{\sci}{Science}

\def\kbar{{\mathchar'26\mkern-9mu k}}
\def\totd{{\mathrm{d}}}
\newcommand{\be}{\begin{equation}}
\newcommand{\ee}{\end{equation}}

%\slugcomment{Accepted for publication in the Astrophysical Journal}

\title{The Intrinsic Two-Dimensional Size of Sagittarius A*}

\author{
 Geoffrey C.\ Bower\altaffilmark{1,2}, Sera Markoff\altaffilmark{3}, 
 Andreas Brunthaler\altaffilmark{4}, Casey Law\altaffilmark{2}, Heino Falcke\altaffilmark{5,4,6},  Dipankar Maitra\altaffilmark{7},
M. Clavel\altaffilmark{8,9}, A. Goldwurm\altaffilmark{8,9},
M.R. Morris\altaffilmark{10}, Gunther Witzel\altaffilmark{10}, Leo Meyer\altaffilmark{10}, and A.M. Ghez\altaffilmark{10}
\altaffiltext{1}{ASIAA, 645 N. A'ohoku Pl., Hilo, HI 96720; gbower@asiaa.sinica.edu.tw}
\altaffiltext{2}{Radio Astronomy Laboratory, UC Berkeley, B-20 Hearst Field Annex, Berkeley, CA 94720-3411}
\altaffiltext{3}{Anton Pannekoek Institute for Astronomy, University of Amsterdam, Science Park 904, 1098XH Amsterdam, The Netherlands}
\altaffiltext{4}{Max-Planck-Institut f\"ur Radioastronomie, Auf dem H\"ugel 69, D-53121 Bonn, Germany}
\altaffiltext{5}{Department of Astrophysics, Institute for Mathematics, Astrophysics and Particle Physics (IMAPP), Radboud University, PO Box 9010, 6500 GL Nijmegen, The Netherlands}
\altaffiltext{6}{ASTRON, P.O. Box 2, 7990 AA Dwingeloo, The Netherlands}
\altaffiltext{7}{Department of Physics and Astronomy, Wheaton College, Norton, MA 02766}
\altaffiltext{8}{AstroParticule et Cosmologie (APC), Universit{\'e} Paris 7 Denis Diderot, 75205 Paris cedex 13, France}
\altaffiltext{9}{Service d'Astrophysique/IRFU/DSM, CEA Saclay, 91191 Gif-sur-Yvette cedex, France}
\altaffiltext{10}{UCLA Division of Astronomy \& Astrophysics, Los Angeles, CA, 90095-1562, USA}
}

\begin{abstract}
We report the detection of the two-dimensional structure of the radio source  associated with the Galactic
Center black hole, Sagittarius A*, obtained from Very Long Baseline Array (VLBA) observations
at a wavelength of 7mm.  The intrinsic source is modeled
as an elliptical Gaussian with major axis size $35.4 \times 12.6\, R_S$ in position
angle 95 deg East of North.  This morphology can be interpreted in the context of both jet and
accretion disk models for the radio emission.  There is supporting evidence in large angular-scale
multi-wavelength observations for both source models for a preferred axis near 95 deg.
We also place a maximum peak-to-peak change of 15\% in the intrinsic major
axis size over five different epochs.
Three observations were triggered by detection of near infrared (NIR)
flares and one was simultaneous with a large X-ray flare detected by NuSTAR.  
The absence of simultaneous and quasi-simultaneous flares indicates that not all
high energy events produce variability at radio wavelengths.  This 
supports the conclusion that NIR and X-ray flares are primarily due to electron
excitation and not to an enhanced accretion rate onto the black hole.
\end{abstract}


\section{Introduction}
\label{sec:intro}

The supermassive black hole at the center of our Galaxy, 
Sagittarius A* (Sgr A*),
was first discovered as a bright, compact radio
core almost forty years ago \citep{1974ApJ...194..265B}.  In the
intervening decades, an enormous and varied body of observational work
has focussed on this source, because although Sgr A* shows only weak
activity, its proximity offers the chance to observe the accretion
flow without confusion from surrounding material \citep[see reviews
in, e.g.,][]{2001ARA&A..39..309M,
2010RvMP...82.3121G,
2010PNAS..107.7196M,
2013CQGra..30x4003F}.

The radio through submillimeter (submm) spectrum of Sgr A* is
inverted, and its polarization, variability and high brightness
temperature all indicate an origin in synchrotron emission.  
The emission has been explained as coming from a jet-like outflow 
\citep{2000A&A...362..113F}
or an optically thin accretion flow with particle acceleration \citep{2003ApJ...598..301Y}.
\citet{2013A&A...559L...3M}
recently presented a combined jet+inflow model based on GRMHD simulations, 
that is able to reproduce the intrinsic size and spectrum of Sgr A* well. 
In that case the major contribution of the size at 7mm wavelength is from the outflow, 
with the disk providing a smaller contribution.  Similar
radio properties in nearby low-luminosity active galactic nuclei
(LLAGN) have been resolved into weak jets (e.g., M81;
\citealt{2000ApJ...532..895B}) but Very Long Baseline Array (VLBA) 
observations have revealed no evidence of an elongated core that would be
indicative of a jet in Sgr A*
\citep{2006ApJ...648L.127B}.  Larger angular-scale imaging in the Galactic
Center, however, has led to several claims for jet-like features
\citep[e.g.,][]{2012ApJ...758L..11Y,2013ApJ...779..154L}.  The inverted spectrum requires only that the
emission is self-absorbed and stratified, i.e., that increasingly
higher frequencies originate in smaller physical scales of the system.
Unfortunately, images at lower frequencies, where more pronounced
elongation would be expected, are blurred out by an intervening
scattering screen \citep[e.g.,][]{2004Sci...304..704,2006ApJ...648L.127B,2007MNRAS.379.1519M}.  
Although the scattering effects
are weak at short wavelengths, structures at these wavelengths
probe regions within $\sim 10$ Schwarzschild
radii ($R_s=2GM/c^2$) from the black hole \citep{2008Natur.455...78D}, where it is difficult to
isolate outflow from inflow.  This problem is
compounded by the fact that the population of particles emitting
synchrotron radiation in quiescence is predominantly quasi-thermal
\citep[see, e.g.,][]{2010PNAS..107.7196M} and, thus, provides a smaller
photospheric ``footprint'' compared to the more extended profile that a
distribution with a nonthermal power-law component would yield.  

Variability provides another tool for probing the structure and physics of Sgr A*.
X-ray flares occur daily in Sgr A* with amplitudes from a factor of a few
to a factor of $>100$; near infrared (NIR) flares appear continuously with on the
order of 4 to 6 peaks per day, often appearing simultaneously with X-ray 
flares \citep[e.g.,][]{2006ApJ...644..198Y,2012ApJ...759...95N}.  
On the other hand, radio and millimeter flux densities vary by 
much less; at 7 mm variations are $\sim 10\%$ rms
on timescales of days
\citep{2004AJ....127.3399H,2006ApJ...641..302M}
while at 1.3 mm variations of 100\% on timescales of $\sim 8$ hours are apparent
\citep{dexter}.  The apparent coupling of NIR and X-ray variability to
radio and millimeter variability with an apparent time delay of hours
should be treated cautiously given the high frequency of events at all wavelengths 
and the inability to obtain continuous coverage over 24 hours at radio wavelengths
\citep[e.g.,][]{2008ApJ...682..373M,2009A&A...500..935E,2011A&A...528A.140T,2012A&A...537A..52E,2012RAA....12..995M}.
The overall weak coupling in amplitude between high and low energies 
suggests that the X-ray and NIR variability may stem from sporadic
episodes of particle acceleration in the inner accretion flow or near
the base of the jets rather than significant changes in the accretion rate
\citep[e.g.,][]{2001A&A...379L..13M,2004ApJ...606..894Y,2004ApJ...611L.101L,2011ApJ...728...37D}.

The intrinsic size of the radio source is a measure of the photosphere, i.e. the surface
at a particular wavelength at which the optical depth is unity.
In the jet model, both more efficient particle acceleration and an increased accretion rate, $\dot{M}$,
will lead to increased jet photosphere length \citep{1993A&A...278L...1F,2007MNRAS.379.1519M}.
\citet{2012ApJ...752L...1M} explore accretion disk size scaling with $\dot{M}$
through theory and simulation and find a broader range of scaling relations depending
on model details.
While the synchrotron cooling time for accelerated particles is
$<1000$ s for typical conditions near the black hole, any
sustained injection or structural changes near the base of the jets,
or disruptions ejected as part of an outflow, could lead to variations
in the jet photosphere shape that would be detectable through imaging techniques.
Depending on the timescales and
characteristics of this variability, such as lag (or lead) time compared
to the NIR/X-ray flare, VLBA observations during flares can help
discern the presence of a jet/outflow, or at the very least better
constrain the geometry and emission processes during flares.

In 2012, we participated in the largest ever multi-wavelength campaign
focusing on Sgr A* to date, arranged around a 3 Megasecond Chandra
HETG ``X-ray Visionary Project'' \footnote{XVP; http://www.sgra-star.com}.
During
this campaign we observed 39 new X-ray flares, roughly tripling the number
observed and allowing for the first statistical analyses
\citep{2012ApJ...759...95N,2013ApJ...774...42N}.  These X-ray flares seem to always be
accompanied by simultaneous NIR flaring, though many NIR flares are
seen without corresponding X-ray activity.  The likely explanation is
that the quiescent X-ray emission is coming from the outer regions of
the accretion flow on much larger scales \citep{2013Sci...341..981W}, and
provides a steady baseline flux that masks smaller nonthermal
flares.   Unfortunately, it is not possible to generate triggers from
Chandra observations on short timescales, but larger NIR flares
($\gtrsim 15-20$ mJy) can be considered thresholds of interesting
activity near the black hole \citep{2011ApJ...728...37D,2012ApJS..203...18W}.
Thus, we chose to trigger 7mm (43 GHz) VLBA
observations based on NIR flares.

The VLBA is able to respond rapidly to external triggers, on timescales as fast 
as $\sim20$ minutes.  Both the VLT and the Keck
Observatory participated in the campaign, however the relative
location of these facilities on the earth compared to the VLBA means
that VLT is in a position to generate fast-response triggers for Sgr
A*, while from Keck one must wait almost a day for Sgr A* to become
visible again, thus probing a much longer timescale.   In this paper,
we describe the results of both types of triggers (\S~\ref{sec:obs}), compare the
observed photosphere to the longterm average based on 
10 observations obtained between 1997 and 2002
\citep[\S~\ref{sec:results}; ][]{2004Sci...304..704}, and discuss
those results in the context of multi-wavelength variability and accretion and 
outflow theory (\S~\ref{sec:discussion}).

\section{Methodology, Observations, and Data Reduction}
\label{sec:obs}

\subsection{Methodology}

The observed radio image of Sgr A* is an elliptical Gaussian with an axial ratio
of $\sim 2:1$ with the major axis in position angle 80 deg East of North.  The major axis size
scales as $\lambda^2$, with a size of $\sim 0.7$ milliarcseconds at 7 mm, which
is strong evidence for the presence of strong interstellar scattering.  The intrinsic size as mentioned
above is defined by the photosphere at the observing wavelength; the observed size is a convolution
of the intrinsic source with the scattering kernel.  Sgr A* 
is heavily resolved by VLBA at this wavelength,  making accurate
measurement of the size challenging.   The orientation of the scattering ellipse is not 
aligned with the Galactic Plane or other known features.  \citet{2014ApJ...780L...2B}
demonstrate that the scattering likely originates at a distance that is several kpc away from
the Galactic Center.

We use closure amplitudes and closure phases to analyze these data
\citep{2004Sci...304..704,2006ApJ...648L.127B}.  The closure amplitude
is defined for stations $ijkl$ as
\begin{equation}
C_{ijkl}={{|V_{ij}| |V_{kl}|} \over {|V_{ik}| |V_{jl}|}},
\end{equation}
where $V_{ij}$ is the complex visibility on baseline $ij$.
Two independent permutations are possible for a set of four stations.  The closure phase
is the sum of visibility phases on a triangle.  The closure amplitude has the property
that it is independent of station-based gain factors.  This independence removes sensitivity to problems
such as elevation- or time-dependent antenna gains due to antenna deformation or changes
in atmospheric opacity.  In principle, amplitude self-calibration can achieve the same
effects but it has limitations in two ways relative to closure amplitude analysis.  First, 
self-calibration can distort the shape of a resolved elliptical Gaussian, especially in
a domain of a small number of baselines and of low elevation observations that are strongly affected by phase decorrelation,
variable atmospheric opacity, and elevation-dependent pointing and antenna gain factors.
Second, error analysis on closure amplitudes effectively takes into
account all possible gain solutions, while error analysis of self-calibrated data is only 
considering one realization of all possible gain solutions.  Thus, closure amplitude modeling
and error analysis is less precise but more accurate than conventional imaging or
visibility analysis of self-calibrated data.

\begin{figure}[p]
\includegraphics{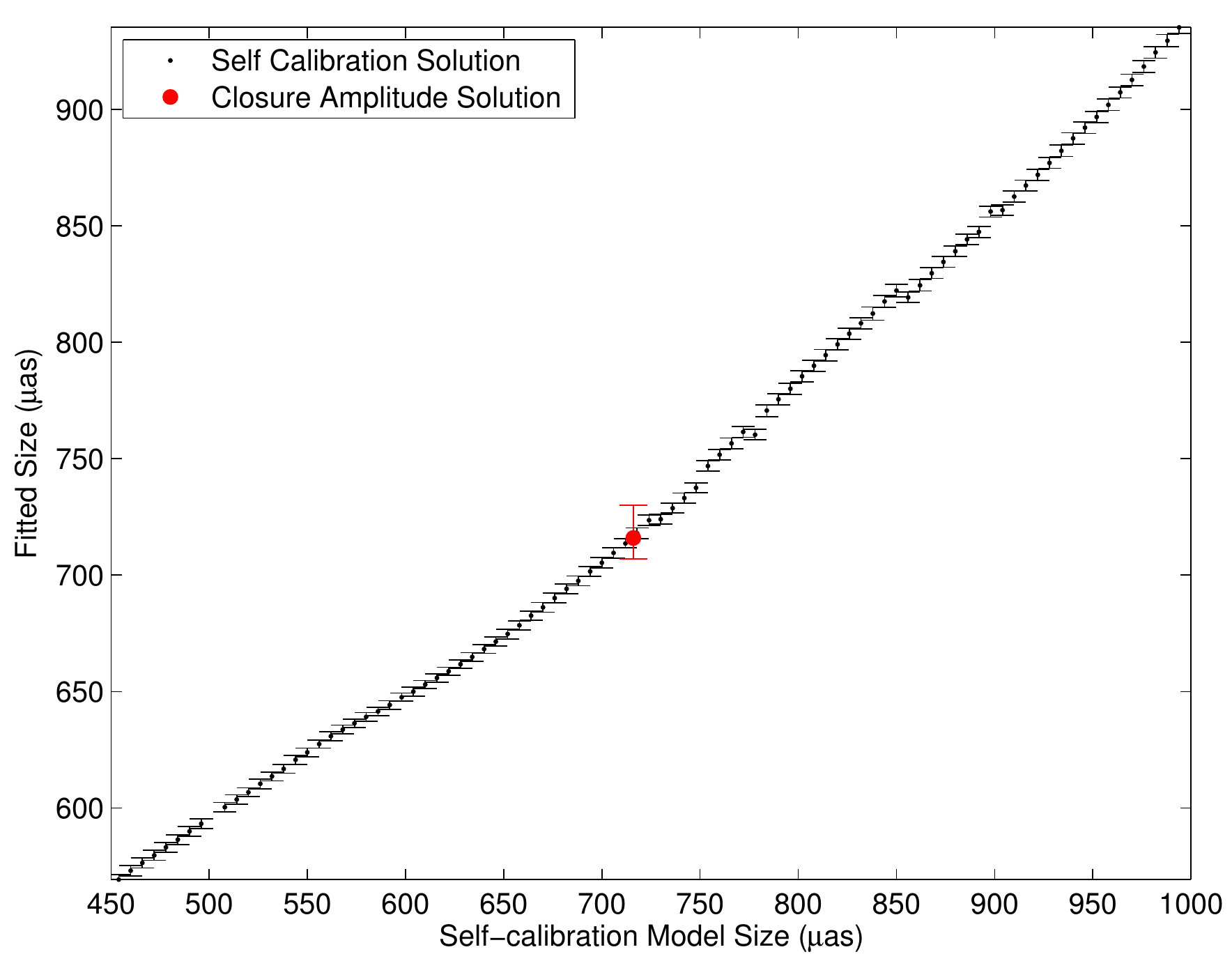}
\caption{
A comparison of self-calbration and closure amplitude techniques for the determination
of the intrinsic size of a resolved Gaussian source.  We compare results
for data obtained on 2012 March 19.  Self-calibration solutions are plotted as a function
of the input self-calibration model.  The closure amplitude solution is also plotted, although
there is no input model offered.  Self-calibration results are heavily dependent on 
the input model and underestimate true errors.  Closure amplitude solutions have
a larger error but provide a single answer that is independent of model assumptions and is,
therefore, more accurate.
\label{fig:selfcaltest}
}
\end{figure}

We conducted a test that demonstrates the limits of amplitude self-calibration clearly
(Fig.~\ref{fig:selfcaltest}).  For the data set obtained on
2012 March 19 (described below), we used amplitude self-calibration and visibility modeling
to determine sizes.  Amplitude self-calibration requires a model, which in this case
we took to be a two-dimensional Gaussian with major axis sizes ranging from 450 to 1000 microarcseconds,
minor axis sizes equal to 50\% of the major axis size, and a position angle of 80 degrees.
The self-calibration algorithm (implemented in AIPS task CALIB) was successful in producing a set of calibrated visibilities.  
The fitted major axis size, however, was strongly dependent on the assumed model:  over
the range of input sizes, the fitted size ranged from 550 to 900 microarcseconds in a roughly
linear trend.  For model sizes less than 420 microarcseconds, the source size was best fit as a
point source.   The median error from the fits was 2.2 microarcseconds, which is significantly smaller
than the closure amplitude errors of $+14$ and $-9$ microarcseconds.  Those self-calibration errors, of course,
misrepresent the true errors in the results because they do not incorporate the errors associated
with antenna gains.  Finally, we note that the self-calibration solution with minimum $\chi^2$ has
a major axis size of $724 \pm 2$ microarcseconds, which is within one $\sigma$ of the closure amplitude solution,
716$^{+14}_{-9}$ microarcseconds.

An alternative approach is to self-calibrate amplitudes on nearby compact sources.  Solutions
obtained from this technique are limited by atmospheric and instrumental fluctuations that can 
give rise to significant errors in amplitudes.

New VLBA observations are significantly more sensitive and accurate than previous
VLBA data for Sgr A*, primarily due to the bandwidth upgrade of the VLBA, which
increased the recording rate from 256 Mb/s to 2048 Mb/s.  
The primary effect of this $8\times$ increase in recording rate is the
ability to detect Sgr A* at high SNR with shorter integration times.  This permits us to remove 
the effects of atmospheric decorrelation with shorter fringe-fitting solution intervals,
thus eliminating a major source of systematic error in closure amplitude analysis.

\subsection{Observations}

\subsubsection{NIR Observations}

In order to select the brightest
flares that are most likely to generate power-law electron tails detectable at radio wavelengths,
we used a trigger threshold of de-reddened NIR flare flux density of 10 mJy.
\citet{2011ApJ...728...37D} and
\citet{2012ApJS..203...18W} analyzed the distribution of $\geq 5$ years of VLT $K_s$-band data.
They found that 
flares with a flux density exceeding 10 mJy occur only $\sim 3\%$ of the time and
so represent the most luminous end of the flare distribution.

The Sgr A* VLT observation that triggered the VLBA observation on 2012 March 19, was part 
of the a multi-wavelength campaign that included a 
120 ks, 5 half-night, X-ray and NIR campaign performed with XMM-Newton and the VLT in order to study the Sgr A*
flaring activity (the detailed results of the campaign will appear in Clavel et al. 2014, in preparation). 
The NIR data were collected using NACO (NAOS+CONICA), the Laser Guide Star and the filter
switching technique already tested in a previous Sgr A* campaign 
\citep{2011A&A...528A.140T},
%%(Trap et al 2011), 
for 5 half nights between 
March 13th and March 21st. These observations were associated with simultaneous observations of Sgr A*
in the range 0.1-10 keV with XMM-Newton.
On March 19th, using the L' filter (at 3.8 $\mu$m) we detected an increase of Sgr A* flux starting around
9:30 UT from an average value of 2 mJy and peaking at about 10 mJy 
(with preliminary peak value of 11.2 +/- 1.1 mJy) at 10:02 UT, when the alert for VLBA triggering was sent. 
In the Ks band (at 2.18 $\mu$m) the measured peak value was about 5 mJy (preliminary peak value of 5.6 +/- 0.4 mJy). 
The given IR flux values are de-reddened, using parameters from \citet{2011ApJ...737...73F}.
Subsequently, the Sgr A* NIR flux started to decrease until the end of the observation run, five minutes later. 
During this period, XMM-Newton was pointed at the target while experiencing stable background and low solar activity,
but did not detect any X-ray flaring from Sgr A*. The XMM observation ended about 10 minutes after the NIR 
flare peak and the telescope was not observing during the VLBA observations that started about 20 minutes
after the NIR peak.

Both telescopes of the W.M. Keck Observatory were used simultaneously during 4 
successive nights: UT 2012 July 20-23 \citep{2014AAS...22323804W}.  Both used laser guide star adaptive optics 
corrections.  The NIRC2 camera was used on the Keck 2 Telescope with an L' (3.8 $\mu$m) 
filter and an observing cadence of 20 seconds.  The imager on the OSIRIS instrument was 
used on the Keck 1 Telescope with an H band (1.6 $\mu$m)  filter and an observing cadence 
of $\sim 50$ seconds.  Poor weather affected the data acquired on July 21 and 22, 
but in any case, no major activity was evidenced on those nights.  On both July 20 and 23, 
however, atypically bright flare events were observed.  
%The L' light curves for those nights are shown in figure X.  
On UT July 20, the flux density of Sgr A* was unusually high throughout the observing 
period with a peak in the L' band of $\sim 18$ mJy at 07:15 UT and continued high flux density
for two more hours. On July 23, 
the flux density rose to a high plateau of $\sim 10$ mJy from a very low value
over the time range 07:30 to the end of the observation at 09:30 UT.
We adopt fiducial flare times of 08:00 and 08:30 UT for comparison with the VLBA
observations on the two successive days, respectively.
The flux densities have been dereddened using the L' extinction from 
\citet{2010A&A...511A..18S}: A$_{L'}$ = 1.23 $\pm$ 0.08.  Stellar calibrators
showed steady flux densities over the observing period.

\subsubsection{VLBA Observations}

VLBA observations were obtained on five epochs (Table~\ref{tab:results}).
Epochs A, B, and C were obtained following NIR triggers (proposal code BM370), while epochs D and E
were scheduled independently of any trigger (proposal code BR173).  All epochs included the inner
six stations of the VLBA (FD, KP, LA, NL, OV, and PT), with the exception of epoch A in which KP was 
missing  due to snow.
The absence of KP significantly reduces sensitivity
to the size because it provides many short and intermediate length baselines.
Data were obtained at a mean frequency of 43.1 GHz with a recording rate
of 2048 Mb/s in both right and left circular polarizations.  Initial data
processing was performed with AIPS \citep{2003ASSL..285..109G}.   Single band
delays were calibrated using observations of bright calibrators including NRAO 530 (J1733-1304)
and J1923-2104. Multi-band delays and fringe rates were determined by
fringe fitting to Sgr A* itself with a 15-second solution interval.  
Longer solution intervals introduce phase decorrelation errors.  The 
antenna-based SNR for fringe solutions on a 15-second timescale was $>10$.
No amplitude calibration was performed on the data.  
Data were averaged in time to a 30-second time interval and over all frequency 
channels.  Data were exported from AIPS and
analyzed with our custom code for closure quantities
\citep{2004Sci...304..704}.

\begin{deluxetable}{llrrrrrr}[p]
%\scriptsize
\tablecaption{Size of Sgr A*\label{tab:results}}
\tablehead{
\colhead{Epoch}&
\colhead{Date}&
\colhead{UT Start}  &
\colhead{UT End}  &
\colhead{$b_{maj}$} &
\colhead{$b_{min}$} &
\colhead{$b_{pa}$} &
\colhead{$\sigma_{int}$} \\
& & (days) & (days) & ($\mu$as) & ($\mu$as) & (deg) & ($R_S$) \\
}
\startdata
  A &  19 March 2012 & 0.431 & 0.677 & $ 716^{+ 16}_{- 11} $ & $ 400^{+ 53}_{-133} $ & $  83.3^{+  3.7} _{-  4.6} $ & $ 33.1^{+ 3.0}_{- 2.7}  $ \\ 
  B &   21 July 2012 & 0.082 & 0.328 & $ 723^{+  3}_{-  3} $ & $ 315^{+ 37}_{- 70} $ & $  83.0^{+  1.0} _{-  1.2} $ & $ 34.7^{+ 0.6}_{- 0.7}  $ \\ 
  C &   24 July 2012 & 0.074 & 0.320 & $ 734^{+  6}_{-  5} $ & $ 300^{+ 61}_{-100} $ & $  81.5^{+  1.6} _{-  2.0} $ & $ 36.9^{+ 1.2}_{- 1.1}  $ \\ 
  D &    13 Feb 2013 & 0.537 & 0.738 & $ 719^{+  4}_{-  4} $ & $ 390^{+ 27}_{- 36} $ & $  81.2^{+  1.2} _{-  1.6} $ & $ 33.7^{+ 0.9}_{- 0.8}  $ \\ 
  E &    23 Feb 2013 & 0.510 & 0.710 & $ 715^{+  5}_{-  4} $ & $ 392^{+ 52}_{-131} $ & $  79.1^{+  2.0} _{-  3.3} $ & $ 32.9^{+ 1.1}_{- 0.9}  $ \\ 
AVG &          \dots & \dots & \dots & $ 722^{+  3}_{-  3} $ & $ 345^{+ 28}_{- 35} $ & $  82.4^{+  0.8} _{-  1.1} $ & $ 34.4^{+ 0.6}_{- 0.6}  $ \\ 
\enddata
\end{deluxetable}

We modeled the size of Sgr A* using closure amplitude data derived from the 
visibilities, as described above.  We explored a range of data selection parameters and found that 
we obtained a minimum in the fitted $\chi^2_\nu$ for the following parameters:
closure amplitude averaging time of 5 minutes; maximum baseline length of 250 M$\lambda$;
and minimum antenna elevation of 10 degrees.  Data were analyzed separately in
right and left circular polarization correlations.  The RCP and LCP results were self-consistent; we averaged
the fits together to produce our final results.  

\begin{figure}[p]
\includegraphics[bb=0.5in 2in 8in 9in,clip=true,width=\textwidth]{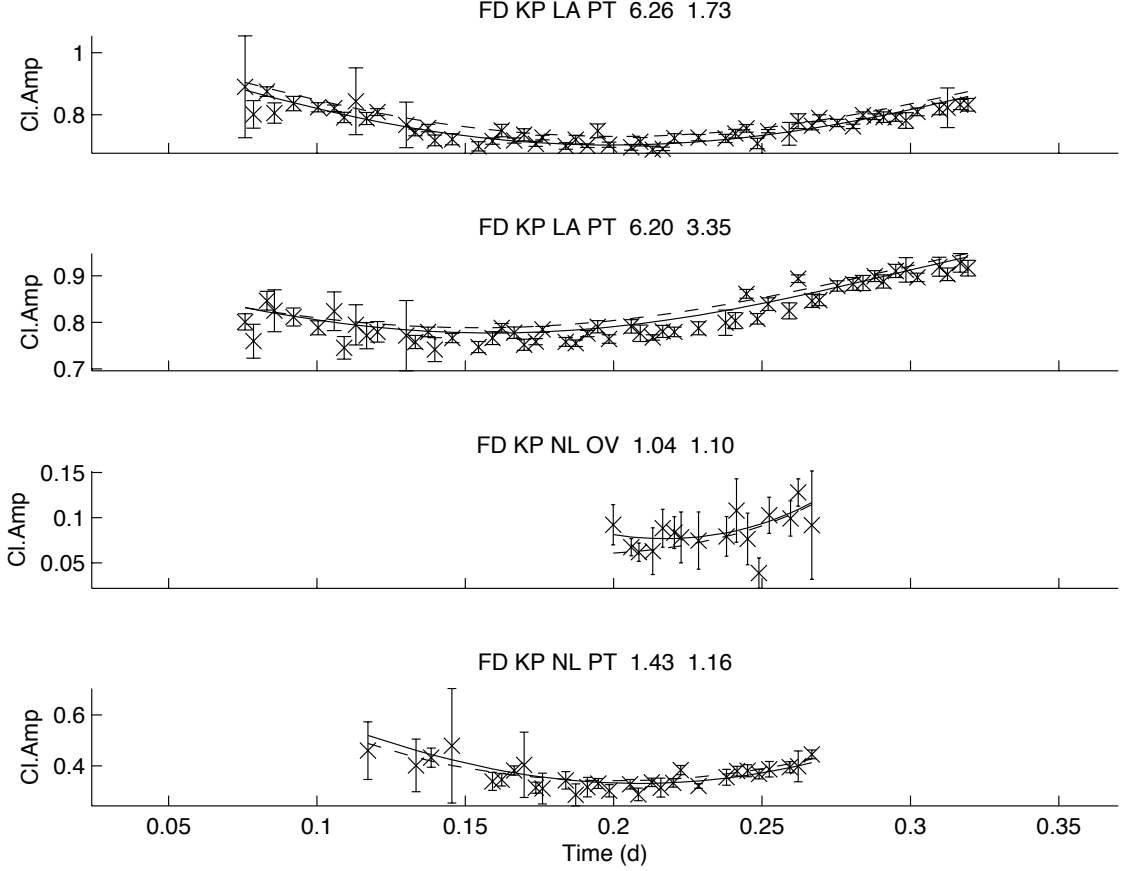}
\caption{
A subset of closure amplitudes for data obtained Epoch C in LCP plotted against UT (in fractional days)
 on 
the day of observation.  For each panel,
the stations involved in the closure amplitude quantity are listed along with the $\chi^2_\nu$
for the average source size and fit source size.  The dashed and solid lines show the 
expected closure amplitudes for the average and fit source sizes, respectively.  
Results are shown for the two independent closure amplitudes generated from the FD-KP-LA-PT
quadrangle.  The FD-KP-NL-OV quadrangle shown involves baselines which exceed the maximum baseline length
(250 M$\lambda$) for our analysis so that only one quadrangle is included.
\label{fig:clamp}
}
\end{figure}

\begin{figure}[p]
\includegraphics{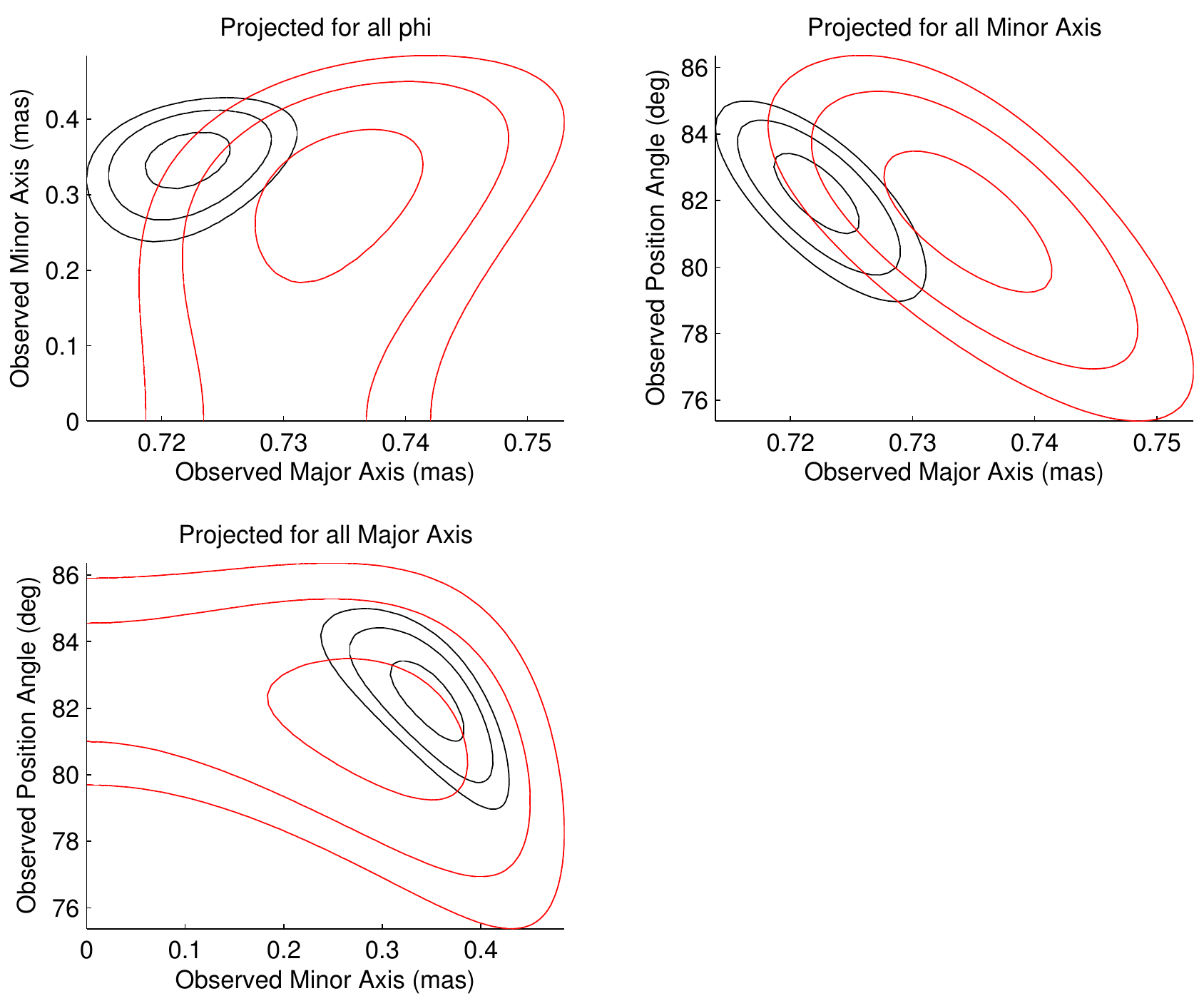}
\caption{Projections of the error surface of the observed size for  Epoch C (red contours) and for the
average of all epochs (black contours).
Panels represent a projection of the three dimensional parameter distribution  
over all position angles, all minor axis values, and all major axis values.
Contours are at 1, 2, and 3 $\sigma$ thresholds.    The figure illustrates the improvement in 
the accuracy with which we measure the parameters through averaging all of the epochs together.
\label{fig:errorsurface}}
\end{figure}

The closure amplitude modeling analyzed the data as a resolved elliptical Gaussian
with parameters:  major axis, $b_{maj}$, minor axis, $b_{min}$, position angle, $b_{pa}$,
the total flux density, $S$, and the noise bias, $s$.  The total flux density is 
not calibrated but is relative to the noise bias, which corresponds to the rms
noise associated with the the 30-second integration time of the visibilities.
We assume that the noise bias $s$ is constant for all antennas and all times,
which is a simplification that limits the number of free parameters.
Representative closure amplitudes are shown in Figure~\ref{fig:clamp}.  We examined 
individual closure amplitude values as a function of time to identify bad
data points and whether there were periods of significant deviation from
the best-fit model.  
A nonlinear least squares method was used to determine the best-fit parameter values.
A grid search of parameters around the best-fit values was used to determine the error surfaces
at $1\sigma$ and $3\sigma$ thresholds.  The error surfaces are not normally 
distributed.  We show projections of the error surface in Figure~\ref{fig:errorsurface}
for one experiment and for the average of all
experiments.   The average of all epochs provides a much stronger constraint
on the parameters than a single epoch.  We take a conservative approach to determining the errors
with two steps:  one, we use the projection of the error surface to determine the
maximum extent of the acceptable parameter value within a confidence threshold; and, two,
we divide the $3\sigma$ error estimates by 3 to determine a
value for the $1\sigma$ error that conveys more information about the bounds of the error 
distribution.  
The minor axis, which is nearly aligned North-South,  is not as 
well constrained as the major
axis due to the predominantly East-West resolution of the VLBA.  In some cases,
the $3\sigma$ error for the minor axis extends to include a value of zero.
Errors for the average are determined by summing together $\chi^2$ surfaces and 
determining the $3\sigma$ surface.
Results from modeling are listed in Table~\ref{tab:results} along with the mean
of all epochs with statistical errors.

We also computed the $\chi^2_\nu$ for the closure phase under the hypothesis
of no axisymmetric structure, for which the closure phase is equal to zero.  
Individual closure phases determined over an averaging time
of 5 minutes had an error of $\sim 1$ deg with peak-to-peak variations of a few degrees.  
For data from Epoch B, we calculated
$\chi^2_\nu = 1.41$ for $\nu=654$.  We find similar results for the other epochs.  Statistically,
this is a large deviation from the null hypothesis but it is also explicable as the result
of a small underestimate of error in the closure phase or the presence of phase
decorrelation.  Zero closure phase is consistent with the model of a resolved elliptical
Gaussian that we use in this paper.  In a future paper, we will explore
detailed comparison with non-axisymmetric models of jets and outflows.

\begin{figure}[p]
\includegraphics{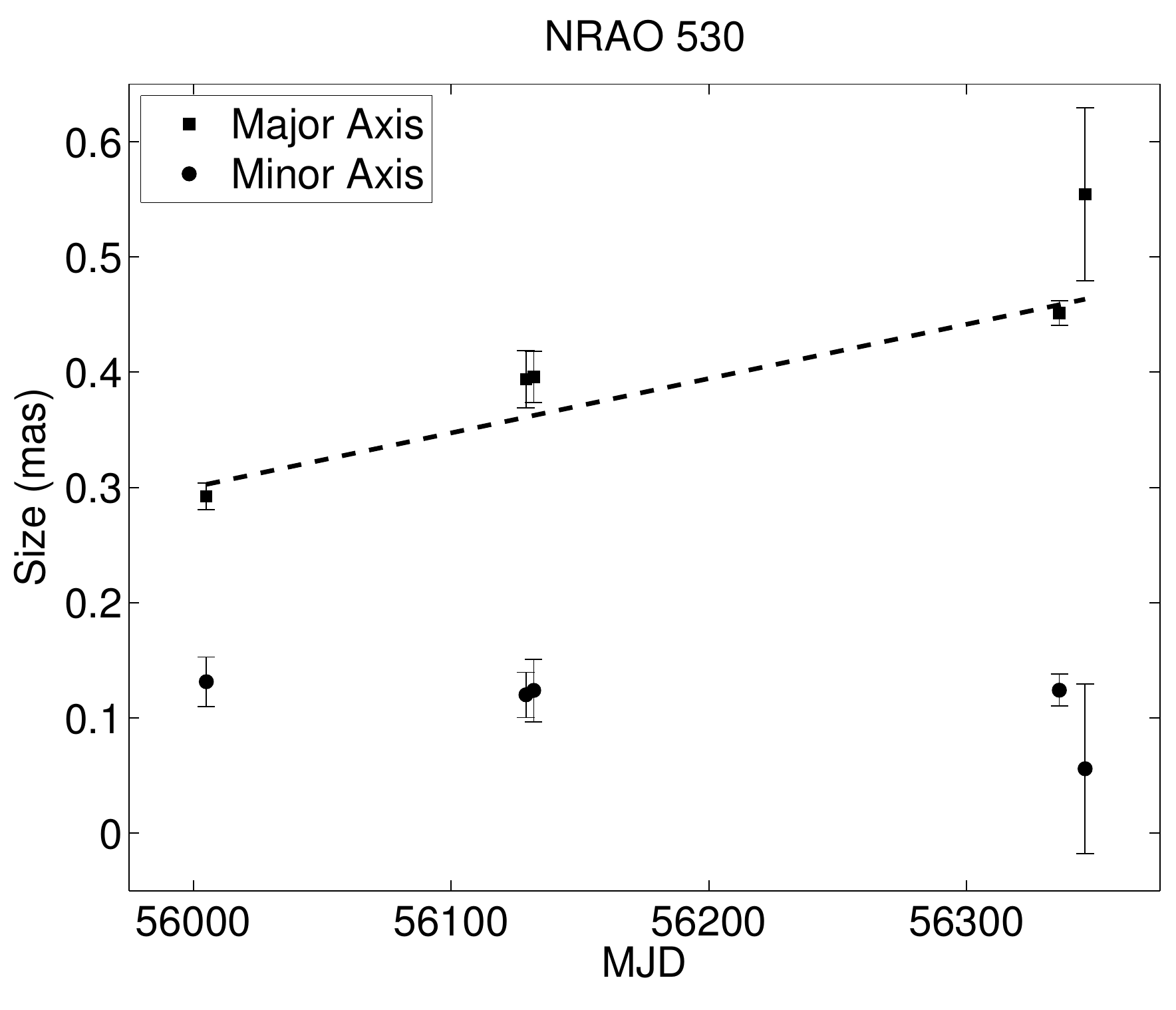}
\caption{Major and minor axis sizes for NRAO 530 as a function of time.
The dashed line indicates a fit to the major axis size that corresponds to an apparent
velocity of $4.3 \pm 0.4c$.
\label{fig:nrao530}
}
\end{figure}

We also applied our technique to the calibrator NRAO 530, which was observed
in all epochs.  The fitted major and minor axis sizes are shown in Figure~\ref{fig:nrao530}.
Sizes are not determined as accurately for NRAO 530 as for Sgr A*, primarily because
of the smaller amount of observing time given to the source.
The minor axis size is consistent with no change over time.  The major axis clearly
changes with time.  This can be fit as an apparent velocity of $4.3 \pm 0.4c$.  The orientation
of the major axis is in position angle $-6 \pm 3$ deg.  Past observations have shown that
NRAO 530 has exhibited outflows oriented North-South with superluminal velocities of $\sim 7c$
at 3mm \citep{1997ApJ...484..118B}, 2 -- $26c$ between 1.3 cm and 3mm \citep{2011MNRAS.418.2260L},  and 6 -- 27$c$ 
at 2cm \citep{2013AJ....146..120L}.  Our measurements are consistent with this historical
evolution.  Unfortunately, the evolution of the source size does not permit us to
determine an absolute standard for stability of the closure amplitude technique with
these data.  But we can place some constraints.  Excluding the final epoch for which data
on NRAO 530 was very limited, the minor axis size shows an rms variation of 5 microarcseconds.
This is an order of magnitude less than the typical error of each measurement, suggesting that we
are overestimating our errors.  The two major axis measurements from 2012 July 21 and 24 
differ by only 2 microarcseconds, while errors are $\sim$ 70 microarcseconds.  This again suggests
that we are overestimating errors in the closure amplitude while achieving substantial accuracy.
Future observations would benefit from inclusion of a true point source that does not evolve.

Closure phases for NRAO 530 also exhibited small deviations from zero indicative of asymmetric structure.
We measured closure phases that deviated from zero by $\sim \pm 5$ deg.  For Epoch B, we calculated
$\chi^2_\nu=6.79$ for $\nu=109$ in marked contrast to the much smaller values obtained for Sgr A*.

\section{The Intrinsic Size}
\label{sec:results}

\subsection{Two-Dimensional Structure in the Average Size}

\begin{figure}[p]
\includegraphics[width=\textwidth]{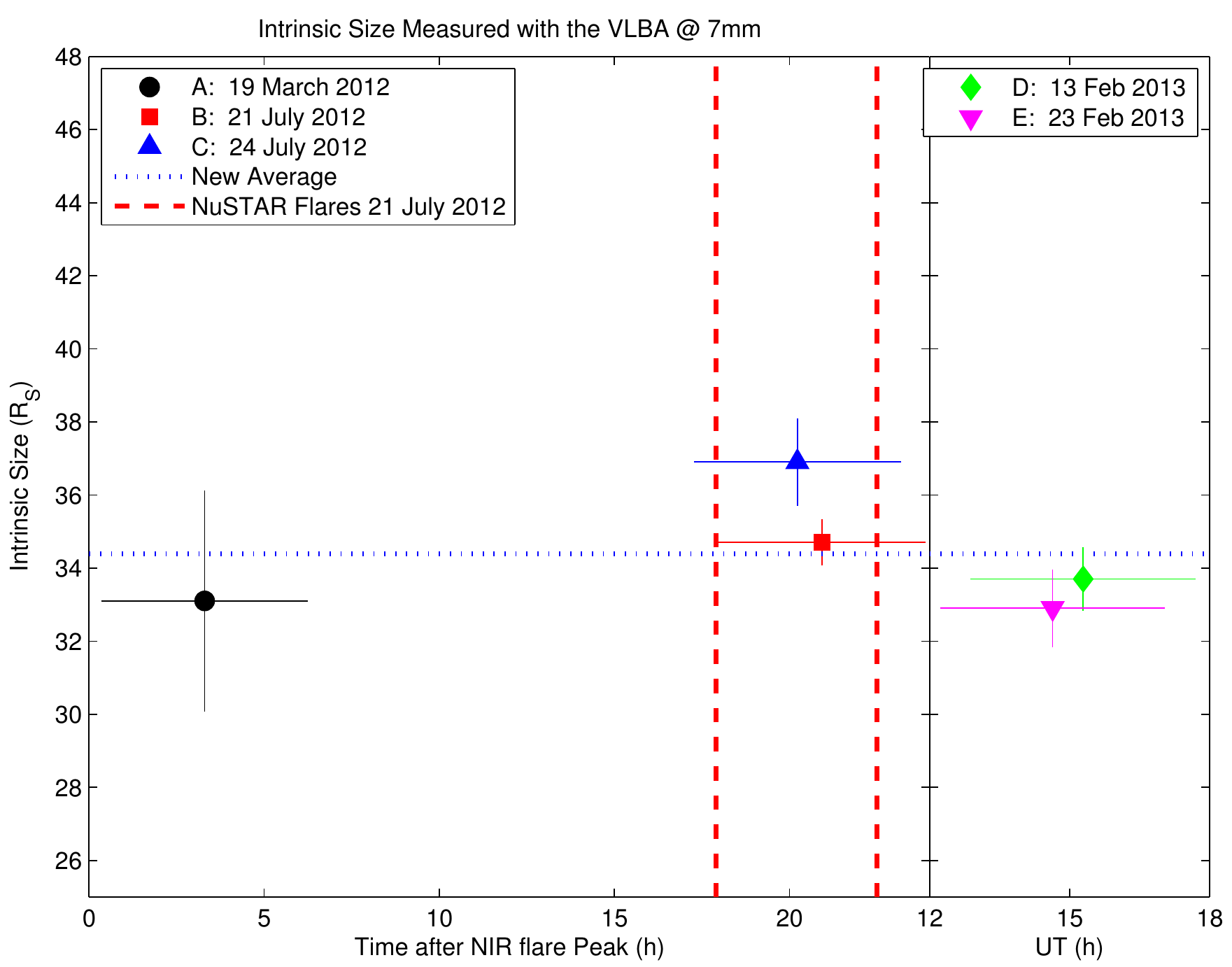}
\caption{
One-dimensional intrinsic size as a function of time delay following the peak of the NIR flare for triggered
observations (left panel) and as a function of UT for non-triggered observations(right panel).  
The horizontal error bars indicate the time range over the which the data was obtained.
The blue dashed line gives the average size from these observations.  Vertical 
dashed red lines give the time of hard X-ray flares detected by NuSTAR on
21 July 2012 (corresponding to Epoch B).
\label{fig:intrinsicsize}
}
\end{figure}

The intrinsic size is computed by subtracting in quadrature the scattering size 
from the observed size.  Longer wavelength observations determine the scattering size
to be an elliptical Gaussian with major axis equal to $(1.31 \pm 0.03) \lambda^2$ milliarcseconds,
minor axis equal to $0.64^{+0.05}_{-0.04} \lambda^2$ milliarcseconds, and position angle equal to $78 \pm 1$ deg,
where $\lambda$ is the wavelength in cm
\citep{2006ApJ...648L.127B}.  Adopting a slightly different major axis scattering scaling constant
as we did in \citet{2009A&A...496...77F}
does not qualitatively alter our conclusions.  At the wavelength of these observations, the mean scattering
ellipse is $635 \times 310$ microarcseconds.  
The best estimate for the Schwarzschild radius based on observational
limits on the black hole mass and distance is 
$1\ R_S=10.2 \pm 0.5 \mu$as plus systematic errors \citep{2010RvMP...82.3121G,2013CQGra..30x4003F}; we adopt a value of 10 $\mu$as for
this paper.  
 Intrinsic sizes in the major axis 
are presented in
Table~\ref{tab:results} and plotted as a function of time delay
following the peak of the NIR flare or the UT time for non-triggered observations
(Fig.~\ref{fig:intrinsicsize}).

\begin{figure}[p]
\includegraphics{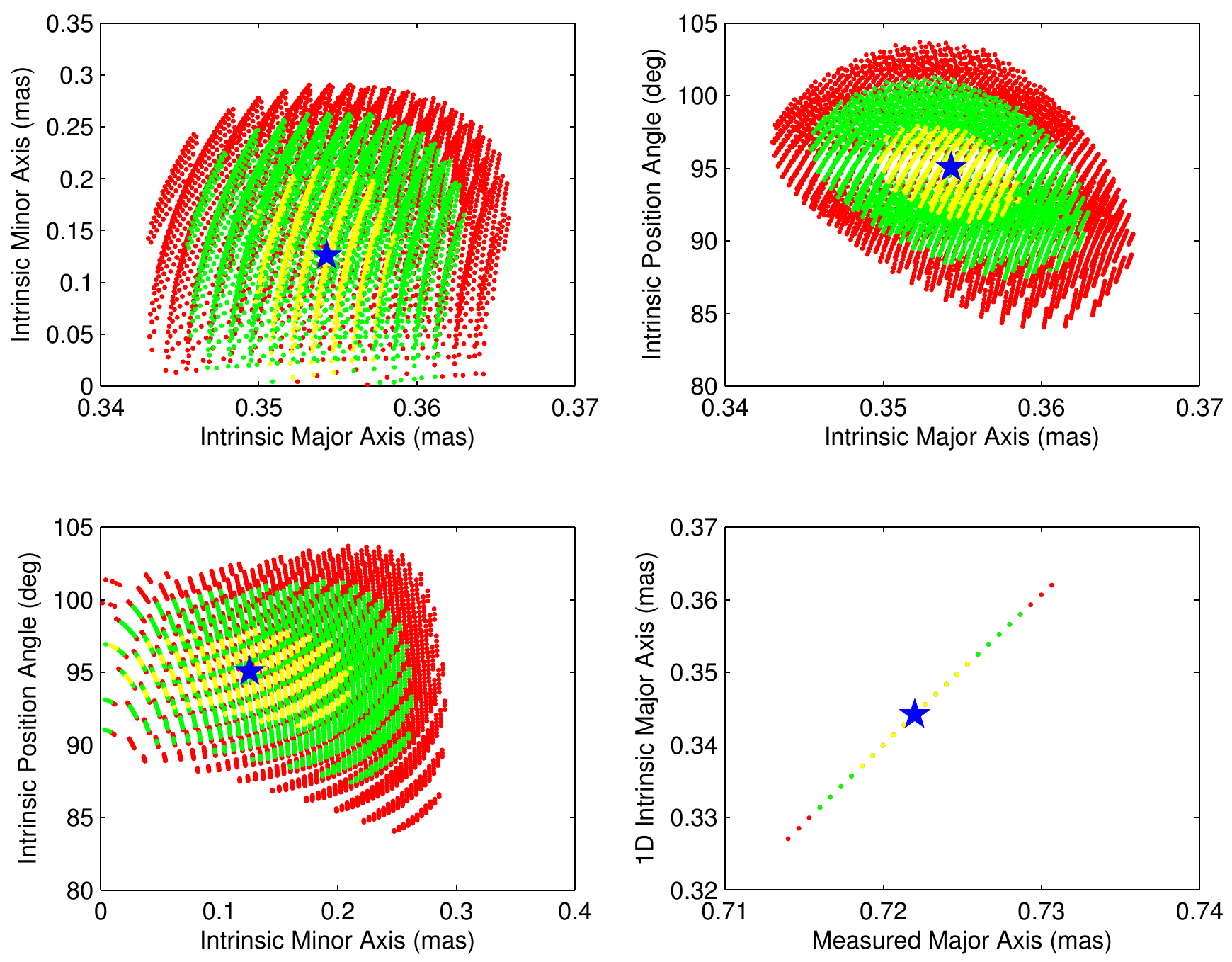}
\caption{Two-dimensional and one-dimensional intrinsic size estimates
for Sgr A*.  Two-dimensional solutions were obtained through deconvolution
of all apparent sizes with the two-dimensional scattering size
within the $3\sigma$ error surface.  Colors show
solutions that are $<1\sigma$ (yellow), between 1 and $2\sigma$ (green), and
between 2 and $3\sigma$ (red).  Slices through the parameter space at
the position of the best-fit position angle, minor axis, and major axis are
shown (upper left, upper right, and lower left, respectively).  The blue 
star indicates the best-fit size.  In the lower right panel, we show the 
intrinsic size that is determined using only the measured major axis size.
The same color coding applies to this panel as to the other three panels.
 The plots show that the maximum intrinsic minor
axis value never reaches the minimum permitted value for the major axis; that
is, the intrinsic source must be elliptical.
\label{fig:intrinsic}}
\end{figure}

We use the average size from all epochs (Table~\ref{tab:results}) to determine a two-dimensional intrinsic
size for Sgr A*.  Deconvolution is performed assuming an intrinsic source and a scattering kernel that 
are two-dimensional Gaussians.
In Figure~\ref{fig:intrinsic}, we show two-dimensional solutions for the intrinsic size for all 
sizes within the $3\sigma$ error surface for which we calculated $\chi^2$ values.
The best-fit size is $35.4 \times 12.6\, R_S$ in position angle 95 deg.  Errors in the major axis size are 
$\pm 0.4 R_S$ and in the minor axis size are $+5.5$ and $-4.1\, R_S$.  
The position angle of the major axis has errors of $+3$ and $-4$ deg. 
The observed minor axis size is only $\sim 1\sigma$ larger than the scattering axis, which effectively
makes the measurement an upper limit on the intrinsic size.
The $3\sigma$ upper limit to the intrinsic minor axis size is $29 R_S$.
If Sgr A* were circular, then the observed minor axis size would be 460 microarcseconds, 
which is more than $4\sigma$ larger than the measured size.  Note that none of
the individual epochs show a minor axis size larger than 400 microarcseconds.
This demonstrates for the first time at any wavelength that the image of Sgr A* is not circular.

Changing the scattering law alters our results quantitatively but not qualitatively.  If we increase
the major axis scattering normalization to 1.36 mas cm$^{-2}$, then the best-fit two-dimensional
solution is $31.3^{+0.4}_{-0.4} \times 11.0^{+5.7}_{-3.5} R_S$ in position angle $101^{+7}_{-4}$ deg.
In this case, a circular source size is rejected with a $3\sigma$ threshold rather than $4\sigma$.

Past measurements of the intrinsic size have only used the observed major axis size.
The mean measured size at the same wavelength from \citet{2004Sci...304..704} is $712
\pm 11\ \mu$as, where the error is based on the scatter in
measurements, corresponding to a size $32.4 \pm 2.5 R_S$.  
If we use the same one-dimensional method to determine the intrinsic size with our
new data, we find an average size of $34.4 \pm 0.6 R_S$.  Note that
the one-dimensional size is slightly smaller than the two-dimensional size.
There is no evidence of change in the average one-dimensional intrinsic size
of Sgr A* between the \citet{2004Sci...304..704} epoch and our new data.
This point is striking given the different nature of the two data sets:  
the older data were not triggered by any measures of activity while the newer 
data track active states of the system, yet there is no difference in intrinsic
size.

\subsection{Variability of the Size between Individual Epochs}

We find no statistically significant evidence for variability in 
the observed major axis, minor axis, and position angle among our five epochs.  
We calculate
$\chi^2_{\nu}=2.0, 0.8,$ and 0.7 with $\nu=4$ for the three parameters, respectively.
While the statistic is larger for the major axis, it does not
demonstrate significant variability.  In Figure~\ref{fig:errorsurface} we can compare
the error surface of Epoch C with that of the average of all epochs.  The mean
major axis size is offset by only $\sim 2\sigma$.

For individual epochs, we use only the observed major axis size to determine a
one-dimensional intrinsic size.  The errors on the minor axis from individual epochs are sufficiently
large that two-dimensional sizes are not well-constrained. 
These one-dimensional sizes are listed in Table~\ref{tab:results} and Fig.~\ref{fig:intrinsicsize}.
We find an rms variation in the intrinsic size of 5\%, with a maximum peak-to-peak change of $\lsim 15\%$.

We note that the \citet{2004Sci...304..704} average size derives from 10 epochs of
observation obtained over 5 years.  The observations were not triggered and so
are likely to represent on average state of Sgr A*.  Given the frequency of NIR flares,
it is not unlikely that the earlier results represent a comparable level of activity
as seen in the new data.  One epoch from \citet{2004Sci...304..704} did
show a formal significant increase in size relative to the mean of $60^{+25}_{-17}$\% (BB130B; 2001 Jul 29),
which was tentatively interpreted as a change in the intrinsic size.  The greater
sensitivity and reduced systematics of this new experiment makes for a
more accurate test of a change in the intrinsic size.  We do not see any variability
with the amplitude seen in the BB130B epoch, although this does not rule out the
possibility that epoch showed actual change in the size.

\citet{2013PASJ...65...91A} presents evidence of structural variability in Sgr A* 
based on VERA observations at 7mm.  They report rms variations in the intrinsic size of 
$\sim 19\%$, which is a factor of several times larger than the rms variability detected here.  These results, however,
are based on traditional calibration and imaging approaches, which suffer from the 
systematic problems discussed in \S~\ref{sec:obs}.
 \citet{2011A&A...525A..76L}  also present limits on variability with a maximum amplitude
of approximately 20\% for 7mm and 3.4 mm observations.
Variability in the size of Sgr A* has not been reported at other wavelengths.  Long wavelength 
sizes appear remarkably stable over decades \citep{2006ApJ...648L.127B}.  Short wavelength sizes have only recently
been determined with high precision.  \citet{2011ApJ...727L..36F} found less than 10\% change in the intrinsic
size at 1.3mm between observations on consecutive days in which the total intensity changed by $\sim$ 40\%.

\section{Discussion and Conclusions}
\label{sec:discussion}

The detection of  elliptical structure and a preferred axis in the geometry of the emitting region of Sgr A*
has important implications.  The axial ratio for the intrinsic size
is 0.35 with a $3\sigma$ upper limit of $<0.85$.
An elongated structure is a possible signature of a jet but could also be
the result of seeing emission from an  elongated accretion flow.  
The position angle of the elongation is primarily East-West ($95^{+3}_{-4}$ deg).  This axis aligns
with estimates of jet orientation based on detailed modeling of the \citet{2004Sci...304..704} 7mm VLBA data along
with spectroscopic constraints, which found
a preferred orientation of 105 deg  with a large uncertainty \citep{2007MNRAS.379.1519M}.   The VLBA orientation does not exactly align with any of
possible jets that have been recently claimed in the literature through arcsecond-scale imaging. 
\citet{2013ApJ...779..154L}, for instance, identify
possible jet interaction with the Sgr A West mini-spiral arm and a linear X-ray feature 
that is directed towards the Southeast of
Sgr A* at a position angle of $\sim 120$ deg.  \citet{2012ApJ...758L..11Y} associate a series of
compact knots and linear features at a position angle of 60 deg.  The apparent jet angle could
differ between small and large angular scales as is often seen in radio-loud extragalactic sources.
If the observed elongation represents an accretion disk, then one might expect
a jet to extend North-South.  This orientation also does not align with either of the putative jet axes.
The East-West orientation also does not align with large scale outflows perpendicular to the Galactic plane
\citep[e.g.,][]{1984Natur.310..568S,2010ApJ...724.1044S}.

On the other hand, X-ray imaging of the quiescent extended emission of Sgr A* has a position angle
of 100 deg and axial ratio of $\sim 0.5$ \citep{2013Sci...341..981W}.  This X-ray elongation
is aligned with the disk of young stars orbiting Sgr A* \citep{2014ApJ...783..131Y}.  The X-ray emission 
represents the accretion flow on the scale of the Bondi radius at a scale of $\sim 10^5 R_S$.
To interpret the radio source as an accretion flow that is coupled 
to the outer scale flow requires that the disk orientation remain the same over linear scales 
that change by four orders of magnitude.  In support of this extrapolation, we note that
\citet{2007A&A...473..707M} do model accretion disk structure on a scale
of a few $R_S$ with NIR polarization data and find a $\chi^2$ minimum at position angle 105 deg albeit
with large uncertainties.  Thus, we find that the VLBA elongation can be viewed as consistent
with evidence for both jet and accretion disk hypotheses.

The detection of  elliptical  structure in our data indicate that shorter wavelength observations
that are less obscured by scattering must explore two-dimensional models in order to
effectively constrain Schwarzschild-scale structure \citep{2008Natur.455...78D}.   Shorter 
wavelength observations currently do not have sensitivity to elongated structures but that
sensitivity will improve as more stations participate in observations. It is uncertain
whether the dominant structures at 7mm will be the same as those at 1.3mm, i.e., there could be a
transition from a jet-dominated to disk-dominated structures between these wavelengths
\citep[e.g.,][]{2013A&A...559L...3M}.  The position
angle may provide a useful prior to constrain models for short wavelength observations that have
very limited coverage of the visibility plane.

We do not find evidence for variations in the size or morphology of the intrinsic size of Sgr A*
in these data or in comparison with historical data.  
The results presented in Fig.~\ref{fig:intrinsicsize} show the intrinsic size of Sgr A*
for the different experiments.  Interestingly, we do not see evidence for a strong
``reaction'' to NIR, X-ray, or millimeter  flares on short ($<$ few
hours) timescales.
The VLT NIR flare (epoch A) had simultaneous
XMM observations that did not detect an X-ray flare.
Both of the Keck triggered NIR flares (epochs B and C) were followed
by X-ray flares detected by Chandra \citep{2013ApJ...774...42N}.  The delays between
NIR and X-ray flares in these cases are hours, suggesting that they may
be unrelated events.  For the case of epoch B, the Chandra observation period
was not simultaneous with the NIR observations.  In the case of epoch C,
the Chandra observation period was simultaneous with the NIR observations but do not
show a simultaneous detected flare.  Epoch B VLBA observations were simultaneous
with NuSTAR observations of a very bright flare \citep{2014arXiv1403.0900B}.   
Millimeter observations with the SMA
appeared to show an increase at 1.3mm wavelength from 3.5 to 4.5 Jy between
9 and 10 UT on 23 July, prior to epoch C (D. Marrone, private communication).  
The NIR flare peaked at 8.5 UT on 23 July.  The 
X-ray flare was detected from 11.8 to 14.0 UT on the same date, beginning
after the SMA observations had ended.  VLBA observations were obtained 
approximately 20 hours after these events.

A change in the photosphere size can result primarily from two processes,
which may or may not be coupled.  In the first case, an increased
accretion rate $\dot{M}$ leads to an increased flux but also increases
the self-absorption frequency.  The exact scaling of the resulting
photosphere size with the accretion rate is, thus, very dependent on the
details of the source geometry and emission processes. For
instance, \citet{1993A&A...278L...1F} predict the
photosphere size to vary $\propto \dot{M}^{2/3}$ for jet models 
while detailed simulations of accretion disks by \citet{2012ApJ...752L...1M}
give a more complex scaling with $\dot{M}$ for accretion disk systems.
If the radio emission is associated with nonthermal particles in the
accretion flow, as suggested by \cite{2003ApJ...598..301Y}, then
an increase in the accretion rate would lead to a size increase in the
radio band first, followed by changes in the submm/NIR/X-ray emission.
On the other hand if the radio emission is due to an outflow/jet, one
would expect the radio photosphere to change only after a rise at NIR and
submm wavelengths.
Thus, changes in accretion rate would produce variability
patterns very dependent on the flow direction, and are thus an important
way to deduce the presence of outflows.   

On the other hand, changes in the radiating
particle distribution will also change the photosphere \citep[see,
e.g.,][]{2010PNAS..107.7196M}, even if there is no change at all in the
accretion rate.  For instance an episode of particle acceleration due
to magnetic reconnection similar to what is seen in the solar corona
would create a power-law component radiating optically thin synchrotron emission
\citep[e.g.][]{2001A&A...379L..13M,2004ApJ...611L.101L,2009ApJ...698..676D,2010AdSpR..45..507T},
and lead to a larger observed size for the duration of the flare.  If this population remains within the inner
parts of the accretion flow, it would not likely affect the
photosphere at 7mm, which corresponds to larger regions in either
the inflow or outflow.  However, if the flaring occurs in plasma
associated with an outflow, it may be carried with the flow on
timescales shorter than the cooling timescales, or reaccelerated, and
thus lead to an observable change.   
A distinguishing characteristic of this model is that these changes
would not have to be associated with a change in the actual flux.   
Unfortunately, the VLBA data do not provide strong constraints on
the source flux density.
Because the cooling timescale for electrons that
create NIR synchrotron emission with conditions near Sgr A*
\citep[$B\sim50-100$ G, $n\sim10^5-10^6$ cm$^{-3}$;][]{2009A&A...508L..13M}
 is less than an hour, the simplest interpretation is that there was a single
episode of particle acceleration that cooled before it could affect
the 7mm band.

Expanding plasmoid models have also been proposed to characterize
the flaring activity of Sgr A* \citep[e.g.,][]{2006ApJ...650..189Y,2006A&A...450..535E,2009MNRAS.395.2183Y}.  
These models attempt to
reproduce the spectral evolution of flares with a spherically expanding
plasmoid of relativistic electrons.  Expansion is typically modeled
as sub-relativistic ($v \lsim 0.01c$) and produces source sizes that are
$\sim 1 R_S$ at NIR and submm wavelengths.  These models have been
shown to require complex and possibly {\em ad hoc} modeling components
to fit with actual flares \citep{2011A&A...528A.140T}.

Our results also limit the degree to which refractive scattering effects can alter the size
of Sgr A*.  The refractive timescale is  $t_{ref} \sim \theta D v^{-1}$, where $\theta$
is the apparent angular size, $D$ is the distance to the scattering medium  from Earth, and $v$ is the relative velocity
of the scattering medium to the Earth.  Recent observations of the Galactic Center pulsar
have shown $D \sim 3$ kpc \citep{2014ApJ...780L...2B}.  Assuming $v \sim 100$ km s$^{-1}$, we find  
$t_{ref} \sim 0.1$ yr at 7 mm.  This sets a minimum timescale in which the scattering medium could change
entirely.  We find no changes to the scattering size on timescales from 2 days to 15 years indicating
that the scattering medium is uniform on scales between $10^{12}$ and $10^{15}$ cm.
Simultaneous shorter wavelength observations could provide an important diagnostic
for separating intrinsic from extrinsic changes in the size.  The relative stability of the size at 1.3 mm
as measured by \citet{2011ApJ...727L..36F} where the refractive timescale is $\sim 1$ day suggest that 
refractive effects are unimportant but they cannot be ruled out.

We have demonstrated two key results with this work.  First, we have measured
the two-dimensional size and orientation of the radio source for the first time.
This provides special axes for exploring large-scale effects that may be associated
with outflows and jets, and accretion flows.  A measurement of the frequency-dependent coreshift
with phase-referenced VLBI observations  is challenging but would confirm the presence and
orientation of a jet.  
Second, we have shown that size variability is very small even in periods
of high activity at NIR and X-ray wavelengths.
This absence of source expansion suggests that these NIR and X-ray flares are driven by a single
particle acceleration event that does not significantly affect the total energy of the
system.  
Denser multi-wavelength observations are essential to understanding the link
between NIR and X-ray variability and the radio source size.

 We demonstrate that the closure amplitude technique is the most accurate method for modeling VLBI data on Sgr A*.
Our results provide an important proof-of-concept and baseline measurement 
for future campaigns on Sgr A* during the upcoming G2 encounter.  
The G2 object is predicted to make a closest approach to Sgr A*
in late 2013 or early 2014 and potentially be disrupted and accreted
onto the black hole
\citep{2012Natur.481...51G,2013ApJ...774...44G,2013ApJ...773L..13P}.
Further tests of accretion, outflow, and emission models will be obtained with
VLBA observations triggered by
increases in the millimeter/submillimeter flux density that are
driven by increases in the accretion rate.

\acknowledgements
The National Radio Astronomy Observatory is a facility of the National
Science Foundation operated under cooperative agreement by Associated
Universities, Inc.   We also acknowledge the role of the Lorentz
Center, Leiden, and the Netherlands Organization for Scientific
Research Vidi Fellowship 639.042.711 for support.   S.M. is also
grateful for support from The European Community's Seventh
Framework Programme (FP7/2007-2013) under grant agreement number ITN
215212 Black Hole Universe.
M.C. and A.G. acknowledge the support from the International Space Science
Institute to the International Team 216 and the financial support
from the UnivEarthS Labex program of Sorbonne Paris Citi{\'e} (ANR-
10-LABX-0023 and ANR-11-IDEX-0005-02). M.C. acknowledges the
Universit{\'e} Paris Sud 11 for financial support.
Parts of this study is based on observations made with ESO Telescopes at the La Silla Paranal 
Observatory under the program VM 088.B-1038 and with telescopes on board XMM-Newton, 
an ESA Science Mission with instruments and contributions directly funded by ESA Member States 
and the USA (NASA). 
A.B. was supported by a Marie Curie Outgoing International Fellowship (FP7) of the European Union (project
number 275596)
H.F. acknowledges funding from an Advanced Grant of the European
Research Council under the European Union’s Seventh
Framework Programme (FP/2007-2013) / ERC Grant
Agreement n. 227610.
The UCLA Galactic center research group acknowledges supported from NSF grant AST-0909218.

%\bibliographystyle{apj}
%\bibliography{myrefs}

\end{document}